\documentclass[aps,twocolumn,groupedaddress,amsmath,amssymb,superscriptaddress,longbibliography]{revtex4-2}
\usepackage[utf8]{inputenc}
\usepackage[T1]{fontenc}
\usepackage[load-configurations=abbreviations, redefine-symbols=true]{siunitx}
\usepackage{upgreek}   
\usepackage{hyperref}
\usepackage{graphicx}  
\usepackage{bm}        
\usepackage{verbatim}   
\usepackage{hyperref}
\usepackage{braket}
\usepackage{lineno}
\usepackage{textcomp}
\usepackage{xcolor}

\begin{document}
\title{Improving the \emph{Q} factor of an optical atomic clock using quantum non-demolition measurement}

\begin{abstract}
Quantum non-demolition (QND) measurement is a remarkable tool for the manipulation of quantum systems. It allows specific information to be extracted while still preserving fragile quantum observables of the system. Here we apply cavity-based QND measurement to an optical lattice clock---a type of atomic clock with unrivalled frequency precision---preserving the quantum coherence of the atoms after readout with 80\% fidelity. We apply this technique to stabilise the phase of an ultrastable laser to a coherent atomic state via a series of repeated QND measurements. We exploit the improved phase-coherence of the ultrastable laser to interrogate a separate optical lattice clock, using a Ramsey spectroscopy time extended from 300~ms to 2~s. With this technique we maintain 95\% contrast and observe a seven-fold increase in the clock's \emph{Q} factor to $1.7\times10^{15}$.
\end{abstract}

\author{William~Bowden}\thanks{these authors contributed equally to this work}
\affiliation{National Physical Laboratory, Hampton Road, Teddington TW11 0LW, United Kingdom}
\author{Alvise~Vianello}\thanks{these authors contributed equally to this work}
\affiliation{National Physical Laboratory, Hampton Road, Teddington TW11 0LW, United Kingdom}
\affiliation{Blackett Laboratory, Imperial College London, Prince Consort Road, London SW7 2AZ, United Kingdom}
\author{Ian~R.~Hill}
\affiliation{National Physical Laboratory, Hampton Road, Teddington TW11 0LW, United Kingdom}
\author{Marco~Schioppo}
\affiliation{National Physical Laboratory, Hampton Road, Teddington TW11 0LW, United Kingdom}
\author{Richard~Hobson}
\affiliation{National Physical Laboratory, Hampton Road, Teddington TW11 0LW, United Kingdom}

\maketitle

\section{Introduction}

In quantum non-demolition (QND) measurement, an observable $\hat{S}$ of a quantum system is coupled to an observable $\hat{M}$ of a \lq meter\rq \, system, so that direct measurement of $\hat{M}$ yields indirect information about $\hat{S}$. While the measurement of $\hat{M}$ may perturb the state of the meter, the inferred value of the observable $\hat{S}$ is conserved by the QND measurement \cite{Braginsky1980}.
QND measurements have given us a window on a wide range of quantum systems, including circuit quantum electrodynamics\cite{Ofek2016,Vool2016,Hacohen-Gourgy2016}, solid-state spin qubits \cite{raha2020, nakajima2019, Xue2020}, mechanical oscillators \cite{Lecocq2015,Rossi2018}, photons \cite{Kono2018,Reiserer2013, Besse2018, Brune1996}, nitrogen-vacancy centres \cite{Cramer2016}, and trapped ions \cite{Hume2007,Wolf2016}.


In this work we use QND measurement to observe cold Sr atoms in an optical lattice clock (OLC), in pursuit of metrological enhancements already demonstrated in Rb- and Cs-based magnetometers \cite{Shah2010, Colangelo2017} and microwave atomic clocks \cite{Hosten2016,Louchet-Chauvet2010, Kuzmich2000,Cox2016,Schleier-Smith2010}. Our work builds on recent demonstrations with Yb \cite{Braverman2019} and Sr \cite{Vallet2017,Norcia2016} by applying QND measurement to a fully operational Sr OLC---an exceptionally stable and accurate type of clock \cite{Oelker2019,Schioppo2017,Ushijima2015,McGrew2018} which is a prime candidate to underpin a future redefinition of the SI second \cite{Lodewyck2019} as well as being a sensitive probe for geodesy \cite{Grotti2018,Takano2016} and physics beyond the Standard Model \cite{Wcislo2018,Delva2017,Takamoto2020,Roberts2020}.

The OLC works by steering the frequency of an ultrastable laser, or \lq local oscillator\rq \, (LO), to match the frequency of the optical $^1$S$_0$ - $^3$P$_0$ clock transition in atomic Sr. The LO frequency is initialised close to resonance with the atomic clock transition, then a spectroscopy pulse is carried out on Sr atoms confined in an optical lattice in the $^1$S$_0$ ground state. At the end of the spectroscopy pulse, the frequency detuning between the LO and the atomic resonance is inferred by measuring the fraction of atoms excited into the $^3$P$_0$ state. In earlier realisations of the OLC \cite{Oelker2019,Schioppo2017,Ushijima2015,McGrew2018} the excitation fraction is measured using fluorescence detection, which destroys the atomic sample. Stabilisation of the LO therefore requires new atomic samples to be prepared, interrogated, and measured in a repeated cycle. By contrast, in this work the excitation fraction is measured using QND methods, allowing the atoms to be recycled for another spectroscopy pulse immediately after measurement.

We carry out QND measurement in an OLC by surrounding the Sr atoms with a high-finesse optical cavity at \SI{461}{\nano\meter}, the wavelength of the strong $^1$S$_0$ - $^1$P$_1$ transition. The same optical cavity also supports a magic-wavelength optical lattice trap \cite{Ye2008}. The \SI{461}{\nano\meter} intra-cavity photons serve as a QND meter of the number of ground state atoms, experiencing a measurable phase shift due to dispersion from the $^1$S$_0$ - $^1$P$_1$ transition.
In this work we demonstrate that, for short probe times, the QND measurement is weak and therefore preserves with high fidelity the coherence of atoms prepared in a superposition of $^1$S$_0$ and $^3$P$_0$. This non-destructive detection enables operation of the OLC in new, more stable configurations, such as
the \lq atom phase lock\rq \ (APL), in which the phase of the LO is stabilised to the phase evolution of the atoms. Here we show that the APL significantly improves the coherence time of the LO laser. Deploying the phase-locked LO in a second, co-interrogated OLC enables us to extend the Ramsey dark time $T$, thereby reducing the Fourier-limited linewidth of the atomic signal $\Delta\nu = 1/(2T)$. This leads to an increased $Q$ factor---i.e. an increased ratio $Q = \nu_0/\Delta\nu$ between the clock transition frequency $\nu_0$ and the spectroscopic linewidth---enhancing a key figure of merit impacting the measurement precision of the clock.

\section{Quantum non-demolition measurement in an optical lattice clock}

\begin{figure*}[tb]
    \centering
    \includegraphics[width=1.4\columnwidth]{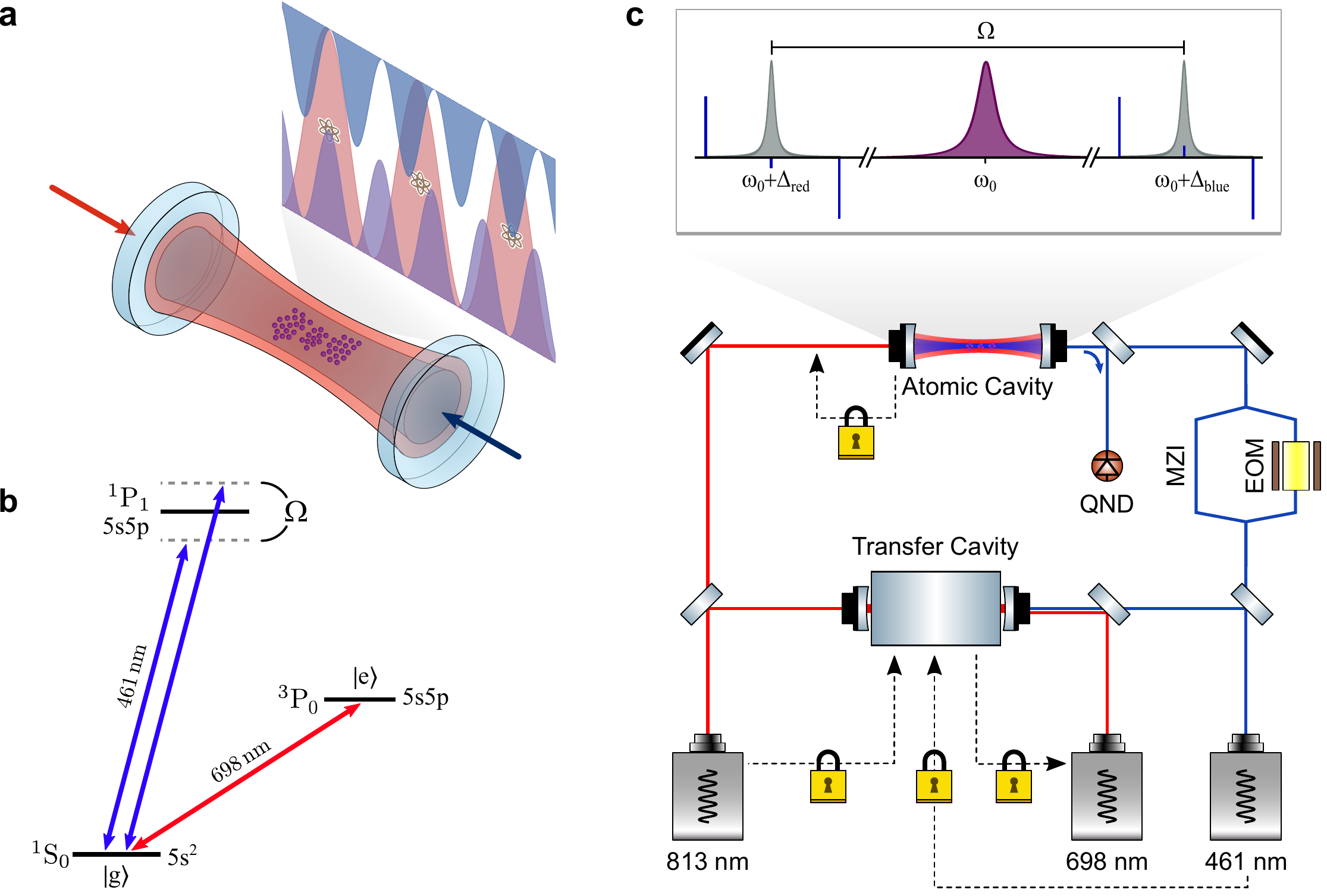}
    \caption{\label{fig:setupForQND} Overview of the QND measurement scheme. \textbf{(a)} Sketch of the dual-wavelength in-vacuum cavity used to trap the atoms and to carry out the QND measurement. Atoms are trapped at the \SI{813}{\nano\meter} intensity maxima represented in red, while they also interact with the nearest blue- and red-detuned cavity modes at \SI{461}{\nano\meter} represented in blue and purple.
    \textbf{(b)} Simplified level scheme for Sr showing the \SI{461}{\nano\meter} transition used for non-destructive detection and the \SI{698}{\nano\meter} optical clock transition. \textbf{(c)} Diagram of the optical setup used for the QND measurement, and a sketch of the optical spectrum transmitted through the Mach-Zehnder interferometer (MZI). The six probe frequency components generated by the electro-optic modulator (EOM) chain are depicted in blue, interacting with the cavity modes in grey which surround the atomic transition in purple. The padlocks represent Pound-Drever-Hall (PDH) loops used to stabilise the laser frequencies and the cavity lengths.
    }
\end{figure*}

To operate the OLC, fermionic strontium ($^{87}$Sr) is laser cooled and loaded into a magic-wavelength, one-dimensional optical lattice at \SI{813}{\nano\meter}. A Ramsey spectroscopy sequence then maps phase or frequency errors of the LO, in this case an ultra-stable laser at \SI{698}{\nano\meter} (see Supplemental Material), onto a population imbalance between the electronic ground state $\ket{g}$ (5s$^2$ $^1$S$_0$, $\mathrm{M}_\mathrm{F} = \pm 5/2$) and the long-lived 
excited state $\ket{e}$ (5s5p~$^3$P$_0$, $\mathrm{M}^{\prime}_\mathrm{F} = \pm 3/2$). Adopting a pseudospin formulation, this population imbalance is encoded in the observable $\hat{S}_z$, the $z$-component of the collective spin of the system $\hat{S}$. The collective spin components can be defined as:

\begin{align}
\hat{S}_{x} = \frac{1}{2}\left(\hat{S}_{\textrm{ge}} + \hat{S}_{\textrm{eg}}\right) \\
\hat{S}_{y} = \frac{1}{2i}\left(\hat{S}_{\textrm{eg}} - \hat{S}_{\textrm{ge}}\right) \\
\hat{S}_{z} = \frac{1}{2}\left(\hat{S}_{\textrm{ee}} - \hat{S}_{\textrm{gg}}\right)
\end{align}
where the operators $\hat{S}_{\textrm{ij}} = \sum_{k = 1}^N \ket{\textrm{i}}_k \bra{\textrm{j}}_k$ are summed over all atoms in the sample.

For a typical OLC, $S_z$ is measured destructively in a two-step process \cite{Oelker2019,Schioppo2017,Ushijima2015,McGrew2018}. First a strong transition at \SI{461}{\nano\meter} from the ground state to an auxiliary state (5s5p~$^1$P$_1$) is used to measure $S_{\textrm{gg}}$ via fluorescence detection. The fluorescence pulse heats the ground state atoms, causing them to escape from the lattice. Next, excited state atoms are optically pumped into the ground state and the fluorescence detection is repeated, giving a measurement of $S_{\textrm{ee}}$. From these two measurements, $S_z$ is calculated and the result is used to correct the LO frequency.

For the OLC in this work, we instead implement a QND measurement of the ground state population using the same optical cavity used to create the one-dimensional lattice trap. The cavity is coated to support optical modes surrounding the \SI{461}{\nano\meter} transition from the ground to the auxiliary state. In the dispersive limit, where the detuning $\Delta$ of the cavity mode from the atomic transition is much larger than the cavity decay rate ($\kappa=2\pi\times$ \SI{330}{\kilo\hertz}), the atomic decay rate ($\Gamma = 2\pi\times$ \SI{30}{\mega\hertz}), and the vacuum Rabi frequency $2g = 2 \pi \times \SI{680}{\kilo\hertz}$, the auxiliary state can be adiabatically eliminated. What remains is an effective coupling between the ground state population $S_{\textrm{gg}}$ and the photon number in the cavity mode $\hat{c}^\dagger \hat{c}$, described by the Hamiltonian \cite{Blais2004, Zueco2009}:

\begin{equation}
\hat{H}_c = \hslash g^2\hat{c}^\dagger \hat{c}\hat{S}_{\textrm{gg}}/\Delta \; .
\end{equation}


This gives rise to an atom-induced frequency-shift of the cavity resonance $\delta\nu = \left<\hat{H}_c/h\right>/\left<\hat{c}^\dagger \hat{c}\right>$. The basic principle of the QND measurement is to drive the cavity with a weak input field at \SI{461}{\nano\meter}, so that the reflected output field carries information about $\delta\nu$, and therefore acts as a meter for the number of atoms. The phase of the reflected field is measured destructively as a beat signal on a photodetector, giving a signal proportional to $S_{\textrm{gg}}$. To obtain $S_z$, which is needed to estimate the LO frequency error, the ground and excited state populations are swapped via a $\pi$-pulse T at \SI{698}{\nano\meter} and a second QND measurement of $S_\mathrm{gg}$ is performed.

Further technical details of the QND measurement scheme \cite{Hobson2019} are outlined in Fig. \ref{fig:setupForQND}. In order to provide first-order immunity to cavity length fluctuations \cite{long2007,Ye1998,Vallet2017}, we probe the difference in the atomic-induced frequency shift between two adjacent longitudinal cavity modes centered in frequency around the atomic transition. 
The optical field used to probe the two cavity modes is generated by sending the \SI{461}{\nano\meter} laser through a Mach-Zehnder interferometer (MZI) amplitude modulator biased to zero throughput and driven at a frequency $\Omega/2 = $ \SI{2.09}{\GHz} matching the \SI{4.18}{\giga\hertz} free spectral range of the cavity. Additional sidebands at $\Omega/2~\pm$ \SI{125}{\mega\hertz} are applied using the MZI modulator, generating strong frequency components which are directly reflected from the cavity input mirror. The strong directly-reflected sidebands interfere with the cavity-coupled probe sidebands at $\pm \Omega/2$, generating a Pound-Drever-Hall \cite{Drever1983} beat signal at \SI{125}{\mega\hertz} proportional to the phase shift induced on the probe sidebands due to the atom-induced cavity shift $\delta\nu$. 

\section{Weak QND measurement and atom coherence preservation}

To a good approximation the value of $S_z$ is conserved after the QND measurement, but other properties of the atomic system can be significantly altered. For example, a fundamental measurement back-action is exerted by photon shot noise in the probe beam, which generates an increase in the uncertainty of $S_y$ as we extract information about $S_z$, in compliance with the uncertainty principle $\Delta S_y \Delta S_z \geq \left<\lvert S_x \rvert\right>/2$. In practice, however, two other technical effects are much larger for the QND scheme in this work: (1) the photon scatter $\Gamma_{\textrm{sc}}$ into free space, and (2) the inhomogenous ac Stark shift $\Delta_{\textrm{ac}}$. Here, we discuss how these two forms of measurement back-action cause decay in the atom coherence $S_x$. We develop a model for the decoherence, and we present experimental data demonstrating weak QND readout of $S_z$ while preserving $S_x$ with 80\% fidelity.

The scatter and the ac Stark shift depend on the radial position $\rho$ and the position $z$ along the cavity axis, according to:
\begin{widetext}
\begin{align}
    \Gamma_{\textrm{sc}}(\rho,z) &= \left<\Gamma_{\textrm{sc}}(0,z)\right>_{z} e^{-\frac{2\rho^2}{w_0^2}}\left[\left(\cos^2 kz + \sin^2 kz \right) + \frac{2\Delta_{\textrm{sum}}\Delta_{\textrm{diff}}}{\Delta_{\textrm{sum}}^2 + \Delta_{\textrm{diff}}^2}\left(\cos^2 kz - \sin^2 kz \right)\right] \label{eq:scatter}\\
    \Delta_{\textrm{ac}}(\rho,z) &= \left<\Gamma_{\textrm{sc}}(0,z)\right>_{z} e^{-\frac{2\rho^2}{w_0^2}} \left[\frac{\Delta_{\textrm{diff}}}{\Gamma}\left(\cos^2 kz - \sin^2 kz \right) + \frac{\Delta_{\textrm{sum}}}{\Gamma}\left(\cos^2 kz + \sin^2 kz \right) \right]\times\left[1 - \frac{2 \Delta_{\textrm{sum}}^2}{\Delta_{\textrm{sum}}^2 + \Delta_{\textrm{diff}}^2}\right] \label{eq:ac_Stark}
\end{align}
\end{widetext}
where $\left<\right>_z$ indicates a spatial average along $z$, $w_0 = \SI{75}{\micro\meter}$ is the waist of the cavity mode, $\Delta_{\textrm{diff}} = (\Delta_{\textrm{blue}} - \Delta_{\textrm{red}})/2 = 2\pi \times \SI{2.09}{\giga\hertz}$ is the average magnitude of the cavity mode detuning, $\Delta_{\textrm{sum}} = (\Delta_{\textrm{blue}} + \Delta_{\textrm{red}})/2 = -2\pi \times \SI{173}{\mega\hertz}$ is the asymmetry of the cavity mode detuning, $\Gamma = 2\pi \times \SI{30}{\mega\hertz}$ is the transition linewidth, and $k = 2\pi/\lambda$ is the wavenumber of the probe. In both equations we have explicitly written separate terms proportional to $\cos^2kz$ and $\sin^2kz$, created by the red- and blue-detuned probe sidebands respectively close to the centre of the optical cavity. Ideally we would simplify the equations by choosing $\Delta_{\textrm{sum}} = 0$, but in practice a small offset is enforced by the technical constraint that the cavity length must be tuned to support a magic-wavelength \SI{813}{\nano\meter} lattice to carry out high-$Q$ spectroscopy on the optical clock transition. Nonetheless we still operate with $ \Delta_{\textrm{diff}} \gg \Delta_{\textrm{sum}}$, such that equation \ref{eq:scatter} yields an approximately uniform photon scatter rate along $z$ while equation \ref{eq:ac_Stark} yields an inhomogenous ac Stark shift varying as $\cos2kz$.

In order to model the effect of $\Delta_{\textrm{ac}}$ and $\Gamma_{\textrm{sc}}$ on the collective atomic spin components $S_i$, we simulate a sample of a few thousand individual spins 
at different positions $\rho,z$ and propagate each spin using optical Bloch equations. The position $\rho$ of each atom is selected from a Gaussian distribution with standard deviation $\sigma_\rho = \SI{35}{\micro\meter}$ corresponding to a radial temperature of \SI{5}{\micro\kelvin}, which has been determined experimentally through sideband spectroscopy \cite{Blatt2009}. Since the radial trap frequency is only \SI{120}{\hertz}, we treat $\rho$ as fixed throughout the QND measurement pulse, which has duration $t < \SI{0.5}{\milli\second}$. The position of each atom along $z$ is randomly selected from one of 2000 sites of the \SI{813}{\nano\meter} lattice trap, matching the experimentally measured width of the cloud. Along $z$, the trap frequency \SI{63}{\kilo\hertz} is comparable to or faster than $1/t$, so we make the approximation that the mean $z$-position of each atom is fixed to the center of the lattice site, but we average the scatter rate and ac Stark shift over a thermal waist $\sigma_z = \SI{50}{\nano\meter}$ corresponding to the \SI{4}{\micro\kelvin} measured axial temperature. 

\begin{figure*}[tb]
    \centering
    \includegraphics[width=1.7\columnwidth]{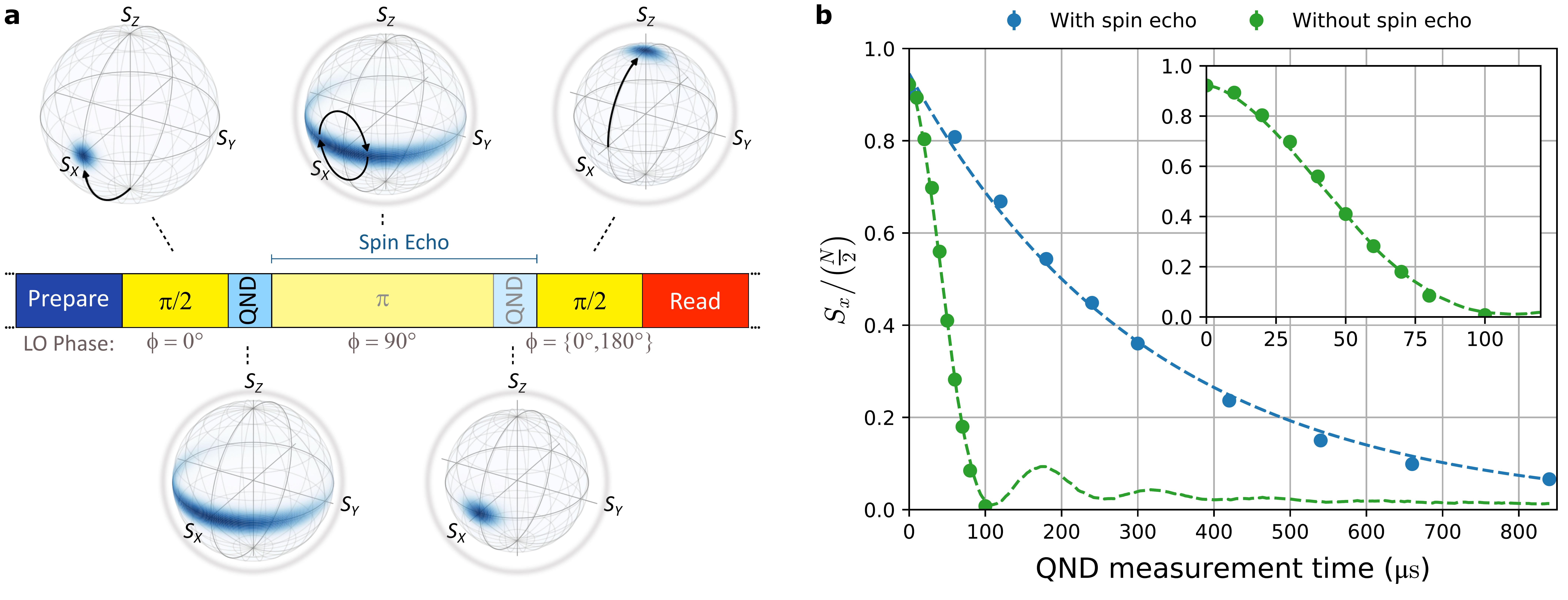}
    \caption{\label{fig:contrast} Coherence preservation after the QND measurement. \textbf{(a)} Timing sequence used to measure coherence after QND measurement, and sketches of the atomic state in the Bloch sphere representation at each step of the sequence. The projection of atoms into the ground or excited state due to scattering of QND probe photons is represented by the shrinking radius of the Bloch sphere compared to its original size $N/2$ (grey halo). The final Bloch sphere (top right) shows the case where the final $\pi/2$-pulse phase is $\phi = 0^\circ$. \textbf{(b)} The measured atomic coherence $S_x$ remaining after QND measurement as a function of total probe time. For the data with spin echo we use the total measurement time summed over the two QND probes, and fit an exponential decay with time constant \SI{317}{\micro\second} (blue dashed line). The model for coherence decay without spin echo (green dashed line, also displayed in the inset) is described in the main text.
    }
\end{figure*}

We investigate the QND probe back-action experimentally using the sequence depicted in Fig. \ref{fig:contrast}a. A sample of \num{6e3} atoms is first prepared in a coherent state with $\left<S_x\right> = N/2$ using a resonant $\pi/2$-pulse from the clock laser. The QND probe is then applied for a variable amount of time $t$. After this, a second $\pi/2$-pulse is applied from the clock laser, the phase of which is stepped by $0^{\circ}$ or 180$^{\circ}$ with respect to the first pulse in order to map $S_x$ to $\pm S_z$. Finally, a destructive measurement is carried out of $S_z$, from which the value of $S_x$ just before the second $\pi/2$-pulse can be inferred. To provide insensitivity to small systematic offsets in the $S_z$ measurement, the estimate of $S_x$ is based on the difference in measured $S_z$ between the two phases $0^{\circ},180^{\circ}$ of the final clock pulse. 
As observed in the data `without spin echo' in Fig. \ref{fig:contrast}b, the inhomogenous ac Stark shift $\Delta_\mathrm{\textrm{ac}}$ results in near-total loss of coherence at QND probe time $t = \SI{100}{\micro\second}$. 
However, the rapid decoherence can be largely reversed using a spin echo protocol. In the `with spin echo' sequence, an additional $\pi$-pulse is inserted with phase $90^{\circ}$ after the first QND probe, followed by a second QND probe. We observe that the decoherence from the ac Stark shift is strongly suppressed by the spin echo, with residual exponential decay of $S_x$ with a time constant \SI{317}{\micro\second} when using \SI{125}{\femto\watt} of cavity-coupled QND probe light. Since the $\pi$-pulse inverts the ground and excited population, the difference between the two QND probe signals in the spin-echo sequence provides a value for $S_z$. Therefore, a spin-echo QND probe sequence with a total probe time $t = \SI{60}{\micro\second}$ can act as a weak measurement of $S_z$, creating a signal to stabilise the clock LO while maintaining coherence with 80$\%$ fidelity.

\section{Increasing the \emph{Q} factor via an atom phase lock}

\begin{figure*}[tb]
    \centering
    \includegraphics[width=1.8\columnwidth]{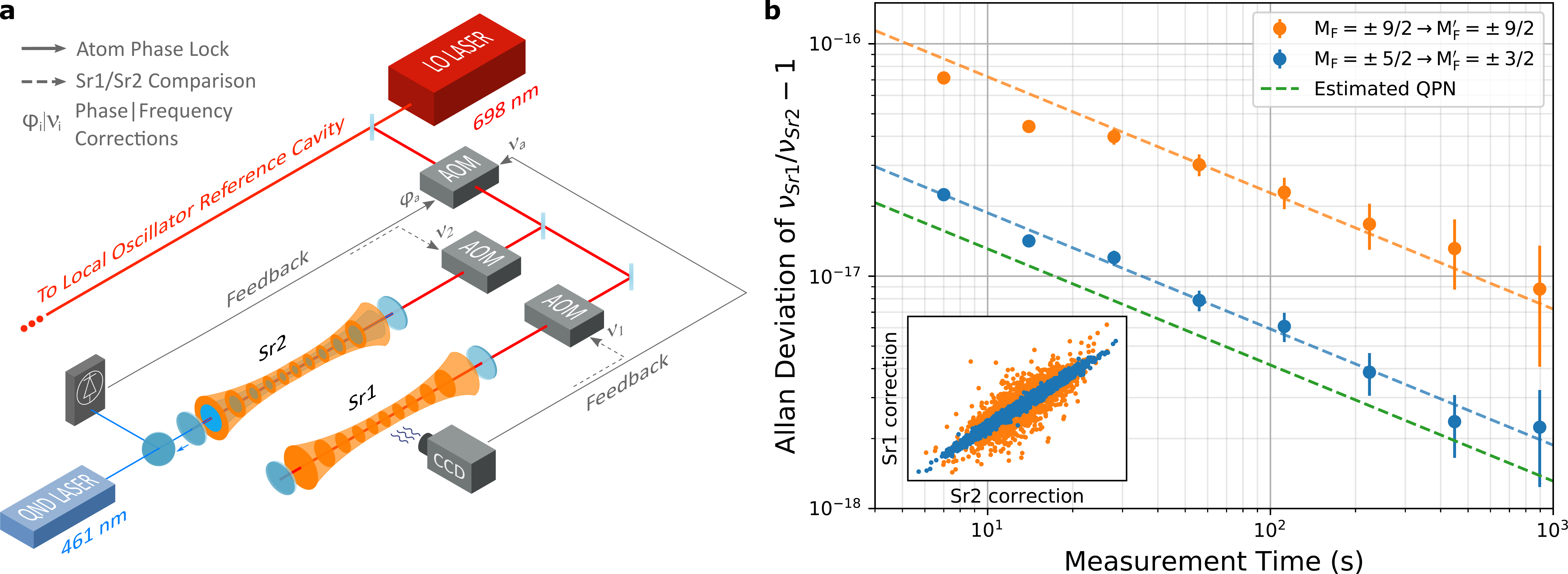}
    \caption{\label{fig:phaseLockSetup} Experimental setup and Sr1/Sr2 correlations. \textbf{(a)} LO light is distributed to Sr1 and Sr2 along separate optical paths, the lengths of which are actively stabilised. When the APL is engaged, phase corrections from Sr2 are applied to an acousto-optic modulator (AOM) shared by both systems, thereby increasing the coherence time of the light sent to Sr1. \textbf{(b)} Allan deviation of the frequency ratio between Sr1 and Sr2, with the APL disengaged and the OLCs independently stabilised using synchronous \SI{300}{\milli\second} Rabi pulses. Both clocks experience the same LO frequency fluctuations, resulting in highly correlated frequency corrections $\nu_1, \nu_2$. However there are residual sources of noise---for example linear Zeeman shift fluctuations, which are suppressed using a less sensitive Zeeman transition. After minimizing noise (see Supplemental Materials), the frequency instability approaches the quadrature sum of QPN from both clocks, $4\times10^{-17}/\sqrt{\tau}$. The cycle time is \SI{1.75}{\second} and atom numbers are \num{7e3} and \num{1.3e4} atoms in Sr1 and Sr2 respectively.}
\end{figure*}

QND measurement in an OLC enables several novel applications that are otherwise impossible using conventional fluorescence readout techniques. Here we pursue one such application---the atom phase lock (APL)---in which the phase noise of the LO is tracked and corrected for via repeated weak measurement of the collective atomic spin. We characterise the performance of the APL to one OLC (NPL Sr2 \cite{Bowden2019a}) using synchronous interrogation of a second OLC (NPL Sr1 \cite{Hill2016,Hobson2020b}) which has highly correlated sensitivity to fluctuations in the LO frequency and phase (see Fig. \ref{fig:phaseLockSetup} and Supplemental Materials).

\begin{figure*}[tb]
    \centering
    \includegraphics[width=1.8\columnwidth]{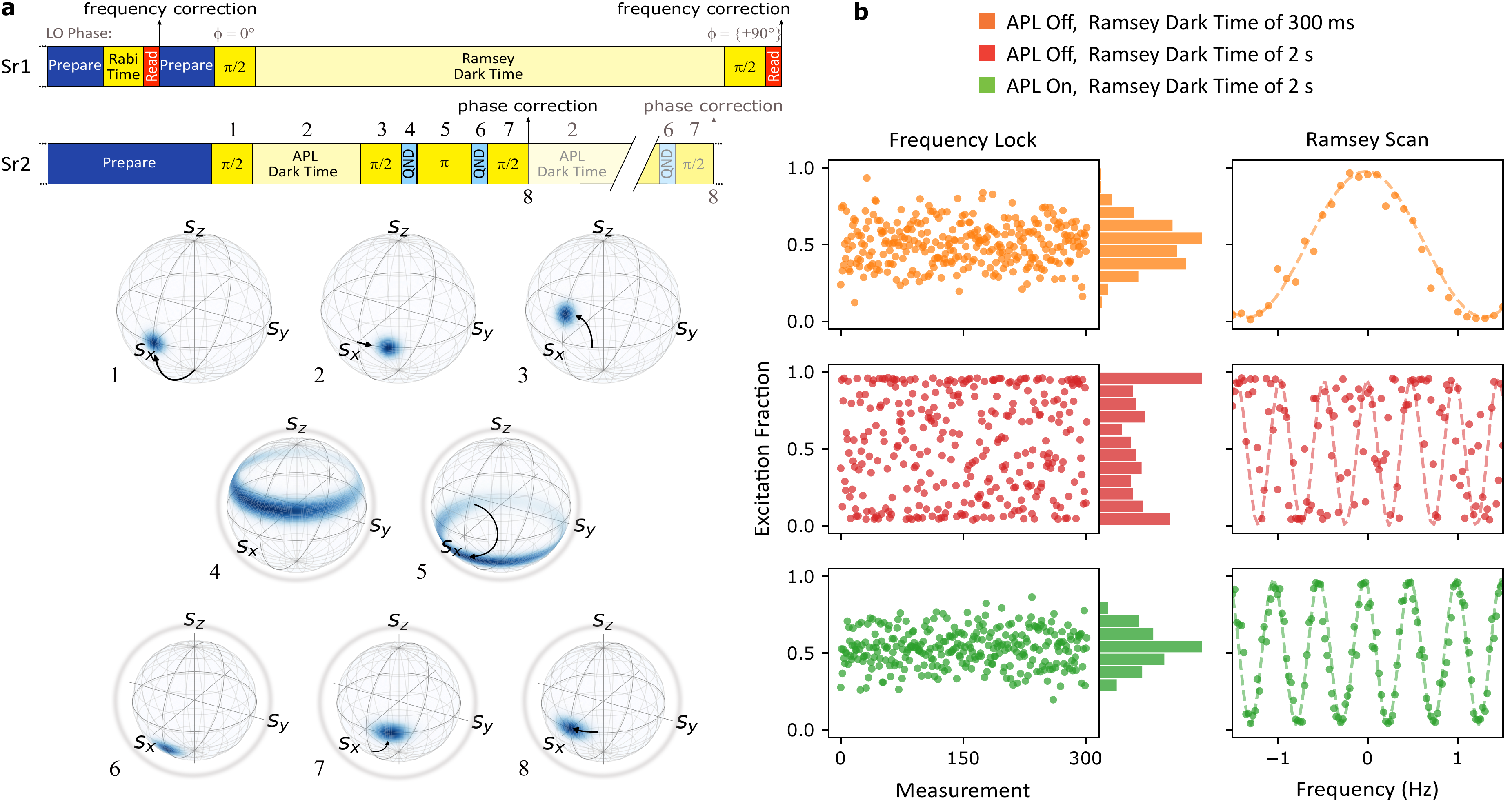}
    \caption{\label{fig:RamseyData} Enhanced Ramsey spectroscopy via APL. \textbf{(a)} Timing sequence for the synchronous spectroscopy scheme, and Bloch sphere representation of the atomic state during the APL sequence. Atomic data from the Rabi time in Sr1 is used only when scanning over Sr1 Ramsey fringes. Prior to and during each scans, the Rabi data measures frequency drift of the free-running LO (typically between zero and \SI{3}{\milli\hertz\per\second}), and allows us to apply LO drift compensation in a double-integrator control loop with attack time of approximately 100 clock cycles. \textbf{(b)} Results using Ramsey spectroscopy in Sr1, co-interrogated with Sr2 using the same LO. Left is the excitation fractions with the LO frequency locked to the central Sr1 Ramsey fringe under three conditions: \SI{300}{\milli\second} Ramsey dark time with the APL to Sr2 disengaged (orange), \SI{2}{\second} dark time with the APL disengaged (red), and \SI{2}{\second} dark time with the APL engaged (green). Frequency scans over the Sr1 Ramsey fringes are shown on the right under the same conditions as above. With the APL engaged, the fringe width is measured to be 254(1)~mHz for a \SI{2}{\second} dark time.
    }
\end{figure*}

After loading approximately \num{1e4} atoms into the optical lattice, the APL is implemented in Sr2 following the scheme depicted in Fig. \ref{fig:RamseyData}{\color{blue}a}, which was originally proposed \cite{Shiga2012} and demonstrated \cite{Kohlhaas2015} for microwave atomic clocks. An initial \SI{10}{\milli\second} $\pi/2$-pulse drives the atomic ensemble into a coherent state on the equator of the Bloch sphere with $\left<S_x\right> = N/2$. As in a normal Ramsey sequence, the atomic state is left to freely evolve during which time it accumulates a phase shift relative to the LO. In the small angle approximation the accumulated LO phase is proportional to $\left<S_y\right>$, which is read out in the following procedure: a $\pi/2$-pulse is driven by the LO, the phase of which is stepped by $90^{\circ}$ with respect to the initial pulse in order to map $S_y$ to $S_z$. The ground-state atom number $S_{\textrm{gg}}$ is then read out via a QND measurement pulse with duration $t = \SI{30}{\micro\second}$. To read out the excited-state atom number $S_{\textrm{ee}}$, a $\pi$-pulse is driven with LO phase $-90^{\circ}$ relative to the initial pulse, before a second QND measurement is applied for $t = \SI{30}{\micro\second}$. 
Finally, the LO phase is stepped again to $90^{\circ}$ and a final $\pi/2$-pulse is applied to return the collective atomic spin to the equator of the Bloch sphere. Based on the results of the two QND measurements, the LO phase is stepped to align the atomic spin to point along the $x$-axis of the Bloch sphere. Repeating the free-evolution time and the QND measurement procedure several times in succession, a phase lock of the LO to the atomic transition can be maintained for several seconds---well beyond the coherence time of the free running LO. 

To characterise the improvement in LO phase noise, the Sr2-phase-stabilised light is used to interrogate Sr1, with results shown in Fig. \ref{fig:RamseyData}{\color{blue}b}. Atomic samples are prepared in parallel in both systems and probed synchronously using the same local oscillator. Sr2 is used to implement the APL while Sr1 performs standard Ramsey spectroscopy. To get a baseline measurement of the free-running laser phase noise, the Sr2 APL is first disengaged and Sr1 is operated as a clock with Ramsey spectroscopy dark time $T = \SI{300}{\milli\second}$. 
When we lock the frequency of the LO to the central Sr1 fringe, we observe noise in the in-lock excitation fraction corresponding to a standard deviation for the accumulated LO phase error of \SI{290}{\milli\radian}. Increasing Ramsey dark time to \SI{2}{\second}, but with the APL still disengaged, shows that the accumulated phase error is too large to operate the clock reliably. This is clear from the U-shaped histogram of the excitation noise in Fig. \ref{fig:RamseyData}, indicating the phase error is well outside $\pm \pi/2$-range that is needed to determine unambiguously the average frequency offset during the dark time. A final dataset is taken with the APL engaged on Sr2 during the \SI{2}{\second} Ramsey dark time in Sr1. Specifically, the APL consists of five repetitions of a \SI{340}{\milli\second} dark time followed by \SI{60}{\milli\second} QND phase measurement and correction. The phase corrections are applied onto an AOM which corrects the LO light prior to it being split and sent to both clocks. Therefore, the average residual phase error accumulated during the APL can be characterised based on the excitation noise in Sr1 when locked to the clock transition, and was determined to be \SI{240}{\milli\radian}. Finally, a scan of the full Ramsey fringe in Sr1 shows no degradation of the 95\% contrast and a Fourier limited linewidth of 254(1) mHz, corresponding to a oscillator $Q$ factor of $1.7\times10^{15}$. This is within a factor of three of the finest scan resolution achieved using state-of-the-art LOs, but unlike earlier high-resolution scans \cite{Campbell2017,Schioppo2017,Norcia2019} we observe no significant loss of contrast on the Ramsey fringes. To our knowledge this matches the narrowest spectroscopic feature to which any oscillator has yet been stabilised \cite{Origlia2018}. Extending the APL time further, either through longer free evolution time or increased number of QND measurements, resulted in increased phase noise in Sr1.

\section{Conclusion}

We have demonstrated that the QND-based APL is an effective approach to improve the phase-coherence of an ultrastable laser, making it a competitive alternative to other strategies for minimising the technical noise of the LO \cite{Matei2017}. Increasing the LO phase coherence directly impacts the frequency stability performance of the OLC, as it enables longer Ramsey dark time $T$, resulting in an increased $Q$ factor and a steeper discriminant of the atomic excitation fraction against the LO frequency. The clearest impact of this is on the quantum projection noise (QPN)-induced fractional frequency instability, which for spectroscopy of $N$ atoms with a signal contrast $C$ and a cycle time of $T_{\textrm{c}}$ is given by:
\begin{equation}
\sigma_{\textrm{QPN}}(\tau) = \frac{1}{\pi QC}\sqrt{\frac{T_{\textrm{c}}}{N\tau}}\;\;,
\end{equation}

\noindent where $\sigma(\tau)$ denotes the Allan deviation for averaging time $\tau$ in seconds. Specifically for Sr1, which operates with \num{5e3} atoms, the seven-fold increase in the $Q$ factor achieved by extending the Ramsey probe time from \SI{300}{\milli\second} to \SI{2}{\second}, with corresponding cycle times \SI{1.3}{\second} and \SI{3}{\second} respectively, reduces the QPN instability from $2.1\times10^{-17}/\sqrt{\tau}$ to $4.8\times10^{-18}/\sqrt{\tau}$.

Another important source of instability in OLCs is the Dick effect, caused by short-term LO frequency noise which is sampled by dead time (primarily cooling time) in the clock sequence. For Ramsey spectroscopy, in the limit of instantaneous $\pi/2$-pulses, the Dick-effect instability is given by \cite{Santarelli1998}:

\begin{equation}
\sigma_{\textrm{Dick}}(\tau)=\sqrt{\frac{1}{\tau}\sum_{k=1}^{\infty}S_{y}(k/T_{\textrm{c}})\left[\frac{\sin(\pi kT/T_{\textrm{c}})}{\pi kT/T_{\textrm{c}}}\right]^{2}}\;\;.
\end{equation}

Increasing the ratio of the Ramsey dark time to the cycle time helps to suppress this effect. Estimating the precise reduction in Dick-effect instability is complex, as it depends on the power spectral density $S_y(f)$ of the fractional frequency fluctuations of the LO at harmonics of the cycle frequency. Under the assumption that LO flicker noise, which we have directly measured to be $8\times10^{-17}$, is the dominant noise process, the estimated Dick-effect induced instability for a \SI{2}{\second} Ramsey dark time is $5\times10^{-17}/\sqrt{\tau}$---a factor of 1.6 below what is expected for a \SI{300}{\milli\second} Ramsey dark time, leading to a reduction in measurement time by a factor of 2.5 to reach the same precision. In the future, an optical frequency comb could transfer the enhanced phased stability of the LO to other wavelengths in order to improve the performance of optical clocks based on different atomic species \cite{Giunta2019, Benkler2019}. In particular, applying this technique to Yb\textsuperscript{+} or highly charged ion clocks which are limited by QPN, but exhibit a large sensitivity to changes in the fine structure constant could facilitate improved tests of fundamental physics \cite{Godun2014, Huntemann2014, Kozlov2018}.



It is instructive to compare our QND-based method against alternative approaches to extend the coherence time of the LO laser. In one demonstration, the LO was pre-stabilised to an OLC with 50\% duty cycle, and then used to interrogate a second OLC \cite{Schioppo2017}. However, with this approach the attainable extension of probe time is limited---there is still considerable dead time of several hundred milliseconds needed to cool atoms in the pre-stabilization OLC, during which the phase of the LO is going unmeasured. Another promising alternative would be to make use of recent advances combining strontium atoms and tweezer arrays \cite{Madjarov2019, Norcia2019}. Such platforms allow for repeated probing of the clock transition and detection, in some cases up to 15 times, without needing to reload the atoms. However, since these experiments rely on fluorescence detection, the phase coherence between the LO and the atoms is lost during detection. If repeated fluorescence readout in a tweezer array were used to implement a destructive form of the atom phase lock, the phase measurement errors (e.g. from quantum projection noise) would accumulate with each interrogation pulse. In contrast, the QND measurement-based approach preserves coherence after each measurement, resulting in phase noise in earlier measurements being corrected for in subsequent ones.

Considering alternative applications of the APL scheme, we speculate that it could help to enable high-Q spectroscopy in environments where the ultimate performance of cavity stabilised lasers can not be reached, for instance in field deployed systems. At the same time, the QND measurement scheme underpinning the APL also opens the door to other configurations of quantum-enhanced optical atomic clocks. Going forward, it will be instructive to characterise the QND measurement scheme in more detail, for example by using colder atomic samples in a better-controlled motional state. Relative to the data presented in Fig. \ref{fig:contrast}, we observe that the QND contrast decay rate can be reduced by a factor of approximately two by adjusting the MZI setup in Fig. \ref{fig:setupForQND} so that the two EOMs generating the stronger probe sidebands at $\Omega/2 \pm \SI{125}{\mega\hertz}$ are placed on different arms of the MZI---this eliminates spurious 2nd-order frequency components near atomic resonance created at the difference frequency between those EOMs, thereby mitigating a source of excess scattered photons. If the QND readout noise and measurement back-action can be controlled close to their shot noise limits, our quantum non-demolition measurement apparatus could be used to generate squeezed states with reduced QPN, offering a route towards OLC comparison with unprecedented frequency precision. Finally, the ability to engineer squeezing in Sr could also have implications beyond precision timekeeping, for example by improving the performance of Sr atom-interferometers \cite{Hu2017} toward the sensitivity necessary to observe gravitational waves \cite{Graham2013}.


\section*{Acknowledgments}

This work was financially supported by the UK Department for Business, Energy and Industrial Strategy as part of the National Measurement System Programme; and by the European Metrology Programme for Innovation and Research (EMPIR) project 17FUN03-USOQS. This project has received funding from the EMPIR programme co-financed by the Participating States and from the European Union’s Horizon 2020 research and innovation programme. A.V. acknowledges funding from the Engineering and Physical Sciences Research Council (EPSRC UK) through the Controlled Quantum Dynamics Centre for Doctoral Training (EP/L016524/1) for the core duration of this work. We thank Helen Margolis, Alissa Silva and Jake Paterson for operating the optical frequency comb, and Rachel Godun and Rich Hendricks for careful reading of the manuscript.

\thispagestyle{empty}

\bibliography{refs}

\section*{Supplementary Materials}

\subsection*{Science packages, local oscillator and stability transfer}

The two OLCs have been described in detail in earlier work (Sr1 \cite{Hill2016,Hobson2020b} and Sr2 \cite{Bowden2019a}). Both clocks run a two-stage cooling sequence, starting with a magneto-optical trap (MOT) operating on the 5s$^2$~$^1\mathrm{S}_0$ - 5s5p~$^1\mathrm{P}_1$ transition. In Sr2, the second-stage MOT is operated on the 5s5p~$^3\mathrm{P}_2$ - 5s4d~$^3\mathrm{D}_3$ transition at \SI{2.92}{\micro\meter}, reaching a temperature of \SI{6}{\micro\kelvin} \cite{Hobson2020}. From the second-stage MOT, atoms load into the optical lattice trap with waist \SI{100}{\micro\meter} and depth \SI{15}{\micro\kelvin}, and after a state preparation and filtering stage \num{1e4} atoms are trapped in the lattice at \SI{4}{\micro\kelvin} and \SI{5}{\micro\kelvin} respectively in the axial and radial directions, prepared into the 5s5p~$^1\mathrm{P}_0$, $\mathrm{M}^{\prime}_\mathrm{F} = \pm 9/2$ state with greater than 99\% purity. In Sr1 the second-stage MOT uses the 5s$^2$~$^1\mathrm{S}_0$ - 5s5p~$^3\mathrm{P}_1$ transition at \SI{689}{\nano\meter}, reaching a temperature of \SI{2}{\micro\kelvin}. From this MOT the atoms load into a vertical, out-of-vacuum cavity-enhanced optical lattice (not yet implemented in \cite{Hill2016}) with waist \SI{150}{\micro\meter} and depth \SI{7}{\micro\kelvin}. After state preparation and filtering, Sr1 has \num{7e3} atoms in the 5s5p~$^1\mathrm{P}_0$, $\mathrm{M}^{\prime}_\mathrm{F} = \pm 9/2$ state trapped in the lattice at \SI{1.1}{\micro\kelvin} and \SI{2.2}{\micro\kelvin} respectively in the axial and radial directions. 

The local oscillator is based on a \SI{1064}{\nano\meter} Nd:YAG laser, frequency stabilised to a reference optical cavity with a length of 485$\,$mm and operated at room temperature \cite{Dovale}. The  fractional frequency flicker floor is measured to be below $8\times10^{-17}$ at \SI{100}{\second} integration time in comparison with a cryogenic laser at Physikalisch-Technische Bundesanstalt (PTB) in Germany through an international optical fibre network \cite{Schioppo2020}. This level of instability is in good agreement with the estimated Brownian thermal noise floor. The stability of the light at \SI{1064}{\nano\meter} is transferred to the Sr lattice clock transition wavelength at \SI{698}{\nano\meter} through a multi-branch frequency comb operated in the transfer oscillator scheme \cite{Telle2002}. Path-length stabilization is implemented to remove phase noise of the optical fibres used to deliver the \SI{1064}{\nano\meter} and \SI{698}{\nano\meter} light to the frequency comb \cite{ma1994}. 

\subsection*{Weak measurement system}

The setup for the QND measurement has previously been presented in detail \cite{Hobson2019} and shares several features with work by another group \cite{Vallet2017}. Since our first demonstration, three major upgrades were made to improve the signal-to-noise ratio as needed to enter the weak measurement regime.

First, a filter cavity was added to remove the amplified spontaneous emission present on the \SI{461}{\nano\meter} probe beam, arising from the \SI{922}{\nano\meter} tapered amplifier which is frequency doubled using a single-pass periodically poled LiNbO\textsubscript{3} waveguide. The \SI{6}{\centi\meter} filter cavity has a finesse of 500, leading to a linewidth of \SI{5}{\mega\hertz}. 
The second upgrade was to replace the fused fibre-optic splitter and the waveguide electro-optic modulator (EOM), originally used in the MZI to generate sidebands at $\Omega/2$ and $\Omega/2 \pm \SI{125}{\mega\hertz}$, 
with a free-space beamsplitter and three dedicated resonant-drive free-space EOMs. This greatly reduced optical losses, since the the waveguide EOM exhibited an unstable insertion loss between 10 and 13~dB. It also eliminated parasitic etalons which had previously compromised the long term stability of the QND signal. Finally, the photodetector used to measure the cavity reflection and create the QND signal, which had an input noise \SI{4}{\pico\ampere\per\hertz\tothe{1/2}} at \SI{75}{\mega\hertz}, was replaced with a resonantly-amplified photodetector with input noise \SI{1.4}{\pico\ampere\per\hertz\tothe{1/2}} at \SI{125}{\mega\hertz} \cite{Bowden2019}, allowing for shot-noise operation at \SI{461}{\nano\meter} for incident power above \SI{18}{\micro\watt}. Together, these improvements enable a detection noise floor within a factor of two of the expected photon shot noise limit in the band \SI{20}{\hertz} - \SI{100}{\kilo\hertz} (excluding harmonics of \SI{50}{\hertz}) for \SI{20}{\micro\watt} total power in the beam incident on the QND photodetector, where previously a factor six excess noise was observed \cite{Hobson2019}.

\subsection*{Minimizing differential phase noise between Sr1 and Sr2}

\begin{figure}[tb]
    \centering
    \includegraphics[width=.95\columnwidth]{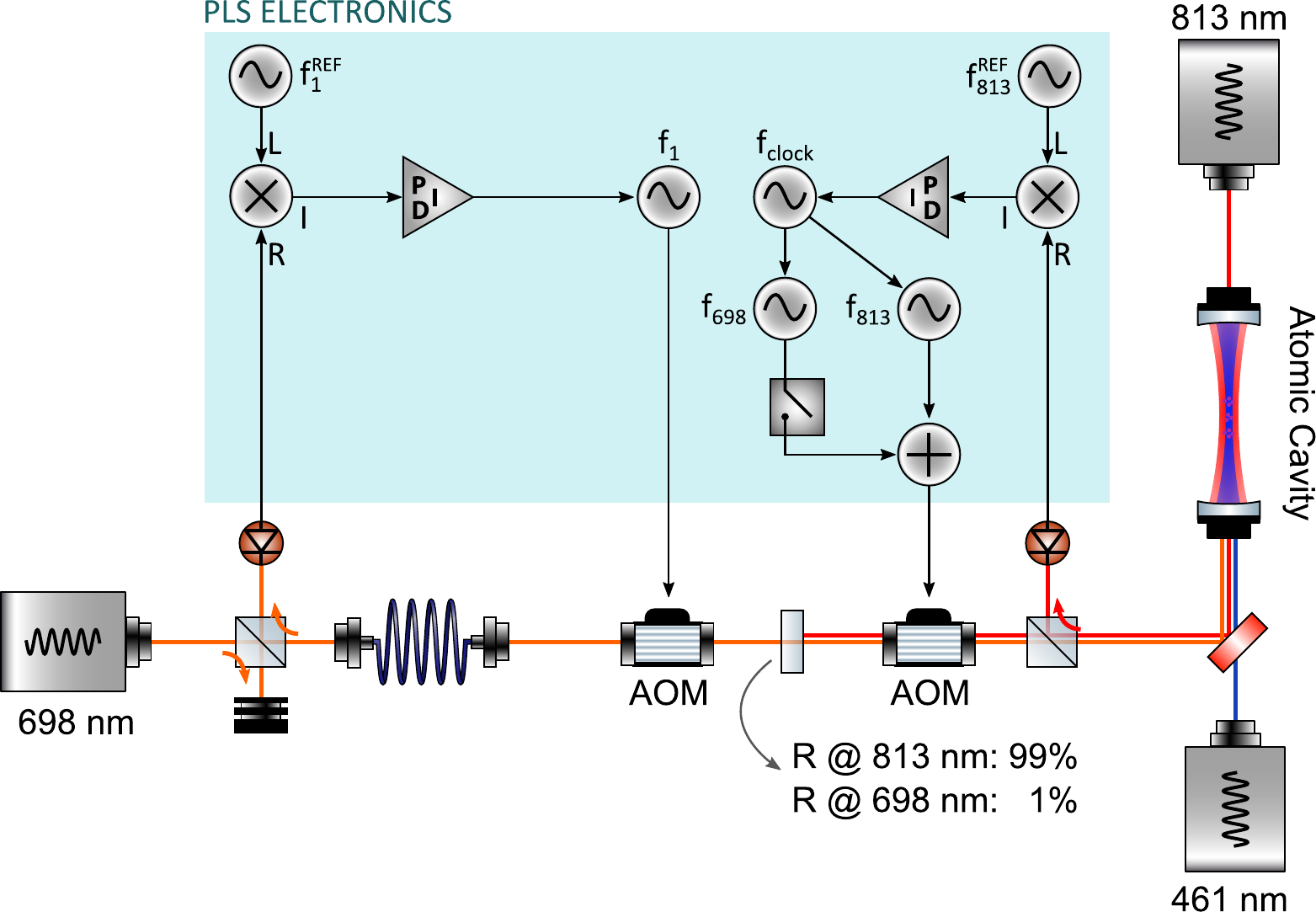}
    \caption{\label{fig:PLS} \textbf{| Path length stabilisation (PLS) scheme.} Optics and electronics underpinning the path-length stabilization for Sr2. Two active path-length stabilization (PLS) control loops are used. The first control loop compensates phase noise in the optical fibre \cite{ma1994}. For the second control loop, the path length detection and feedback is performed using the \SI{813}{\nano\meter} light transmitted through the atomic cavity, and path length corrections are fed forward onto the \SI{698}{\nano\meter} probe light. The AOM frequencies are set to a ratio $f_{698}/f_{813} = \lambda_{813}/\lambda_{698}$ corresponding to the ratio of the optical carrier wavelengths, so that both beams are deflected by the same angle and acquire the same fractional frequency corrections in the phase stabilisation loop. An additional copy of the same system is in place to phase stabilise the light delivered to Sr1, except that the \SI{461}{\nano\meter} QND probe laser is not present.}
\end{figure}

In order for Sr1 to benefit from the improved LO coherence enabled by the Sr2 APL, differential phase noise between the two systems must be controlled to a negligible level. Here we describe two causes of phase noise, and our approach to suppress them.

The largest source of differential phase noise is due to fluctuations in the optical path length traversed by the LO light as it is delivered from the laser to the two atomic systems. Both Sr1 and Sr2 use the same architecture to suppress the path length instability, depicted in Fig. \ref{fig:PLS}. The LO laser light is delivered by optical fibre to each science chamber before passing through an acousto-optic modulator (AOM). The light is then partially retro-reflected back to the LO laser through the AOM and fibre, and is phase locked to a common reference beam via optical heterodyne detection feeding back to the frequency of the AOM drive \cite{ma1994}. In order to stabilise the final $\sim \SI{0.5}{\meter}$ free-space path from the partial retro-reflector to the lattice mirror, we implement another active stabilization loop. In this case, the AOM deflecting the \SI{698}{\nano\meter} LO beam must be switched on and off to implement the Ramsey spectroscopy, thereby interrupting any beat signal at \SI{698}{\nano\meter} which might otherwise have been used for path length stabilization. To solve this problem, we continuously drive the switching AOM with a second RF frequency to diffract the \SI{813}{\nano\meter} lattice light transmitted through the cavity. The lattice light is then reflected by the partial retro-reflector, diffracted by the switching AOM a second time, before returning to the cavity where its phase is compared to the field directly exiting the cavity via a heterodyne measurement. The beat is stabilised via feedback to the clock reference used to synthesise both RF drive frequencies deflecting \SI{698}{\nano\meter} and \SI{813}{\nano\meter}. By choosing the ratio of the \SI{698}{\nano\meter} and \SI{813}{\nano\meter} RF drive frequencies to match the ratio of their optical frequencies, both beams are deflected by the same angle and the path length corrections acting on \SI{813}{\nano\meter} are fed forward with the correct amplitude onto the \SI{698}{\nano\meter} light.

A second source of differential phase noise between the Sr atoms in Sr1 and Sr2 arises from temporal fluctuations in environmental parameters perturbing the clock transition. In order to detect this noise, we carry out synchronous interrogation of the clock transition in both Sr1 and Sr2 using \SI{300}{\milli\second} Rabi spectroscopy from the same LO. In the case of zero differential phase noise, the LO frequency noise-induced fluctuations in excitation fraction in the two systems should be perfectly correlated. Therefore, once we lock the frequency of the LO to Sr1 and Sr2 using independent AOMs with matched servo gains, the frequency instability of the Sr1/Sr2 ratio should be limited only by detection and QPN noise \cite{Schioppo2017}. However, if differential phase noise is present then the excitation fractions will be less correlated and a residual Dick-effect instability \cite{Quessada2003} will arise in the synchronous Sr1/Sr2 ratio. In our system, an important source of differential noise was eventually found to be magnetic field noise in the vicinity of Sr2, which initially limited the Sr1/Sr2 ratio stability to a level a factor of five above the expected QPN limit. In these initial measurements we used the 5s$^2~^1\mathrm{S}_0,~\mathrm{M}_\mathrm{F} = \pm 9/2$ to 5s5p~$^3\mathrm{P}_0,~\mathrm{M}^{\prime}_\mathrm{F} = \pm 9/2$ transition with linear Zeeman sensitivity \SI{4.9}{\kilo\hertz\per\milli\tesla}, due to the relative ease of optically pumping a large fraction of the atoms into the stretched states $\mathrm{M}_\mathrm{F} = \pm 9/2$. However, a factor of four improved instability was observed by instead operating Sr2 on the 5s$^2~^1\mathrm{S}_0,~\mathrm{M}_\mathrm{F} = \pm 5/2$ to 5s5p~$^3\mathrm{P}_0,~\mathrm{M}^{\prime}_\mathrm{F} = \pm 3/2$ transition, which has much lower sensitivity of \SI{280}{\hertz\per\milli\tesla}. The atoms are prepared in this state by first optically pumping to a stretched state, before coherently transferring the atoms using multiple clock pulses to drive $\mathrm{M}_\mathrm{F} = \pm 9/2 \rightarrow \mathrm{M}^{\prime}_\mathrm{F} = \pm 7/2 \rightarrow \mathrm{M}_\mathrm{F} = \pm 5/2 \rightarrow \mathrm{M}^{\prime}_\mathrm{F} = \pm 3/2$. After the coherent transfer, residual atoms remaining in the ground state are cleared out of the lattice using a pulse of \SI{461}{\nano\meter} light. Using the less sensitive Zeeman transition, and keeping all path-length stabilisation loops engaged, we observe a correlation of $R^2 = 0.98$ between the excitation fractions and an instability in the Sr1/Sr2 frequency ratio of $6\times10^{-17}/\sqrt{\tau}$, marginally above the estimated instability arising from the combined QPN of both systems of $4\times10^{-17}/\sqrt{\tau}$.

\end{document}